\documentclass[sigconf]{acmart}
\AtBeginDocument{%
  }
\setcopyright{acmcopyright}
\copyrightyear{2023}
\acmYear{2023}
\usepackage{amsmath,amsfonts}
\usepackage{subfig}
\usepackage{graphicx}
\usepackage{textcomp}
\usepackage{xcolor}
\usepackage[ruled, vlined]{algorithm2e}
\usepackage{algpseudocode}
\usepackage{cuted}
\usepackage{url}
\usepackage{pifont}
\newcommand{\TN}{M2V}
\begin{document}
\title{Having Difficulty Understanding Manuals? Automatically Converting User Manuals into Instructional Videos}
\author{Songsong Liu}
\email{sliu23@gmu.edu}
\affiliation{%
  \institution{George Mason University}
  \city{Fairfax}
  \state{VA}
  \country{USA}
}
\author{Shu Wang}
\email{swang47@gmu.edu}
\affiliation{%
  \institution{George Mason University}
  \city{Fairfax}
  \state{VA}
  \country{USA}
}
\author{Kun Sun}
\email{ksun3@gmu.edu}
\affiliation{%
  \institution{George Mason University}
  \city{Fairfax}
  \state{VA}
  \country{USA}
}
\begin{abstract}

While users tend to perceive instructional videos as an experience rather than a lesson with a set of instructions, instructional videos are more effective and appealing than textual user manuals and eliminate the ambiguity in text-based descriptions. However, most software vendors only offer document manuals that describe how to install and use their software, leading burden for non-professionals to comprehend the instructions. In this paper, we present a framework called \TN{} to generate instructional videos automatically based on the provided instructions and images in user manuals. \TN{} is a two-step framework. First, an action sequence is extracted from the given user manual via natural language processing and computer vision techniques. Second, \TN{} operates the software sequentially based on the extracted actions; meanwhile, the operation procedure is recorded into an instructional video. We evaluate the usability of automatically generated instructional videos via user studies and an online survey. The evaluation results show, with our toolkit, the generated instructional videos can better assist non-professional end users with the software operations. Moreover, more than 85\% of survey participants prefer to use the instructional videos rather than the original user manuals.
\end{abstract}

\begin{CCSXML}
<ccs2012>
   <concept>
       <concept_id>10003120.10003145</concept_id>
       <concept_desc>Human-centered computing~Visualization</concept_desc>
       <concept_significance>500</concept_significance>
       </concept>
   <concept>
       <concept_id>10010147.10010341</concept_id>
       <concept_desc>Computing methodologies~Modeling and simulation</concept_desc>
       <concept_significance>500</concept_significance>
       </concept>
 </ccs2012>
\end{CCSXML}

\ccsdesc[500]{Human-centered computing~Visualization}
\ccsdesc[500]{Computing methodologies~Modeling and simulation}

\keywords{User Manual, Instructional Video Generation, Action Emulation}

\begin{teaserfigure}
  \includegraphics[width=\textwidth]{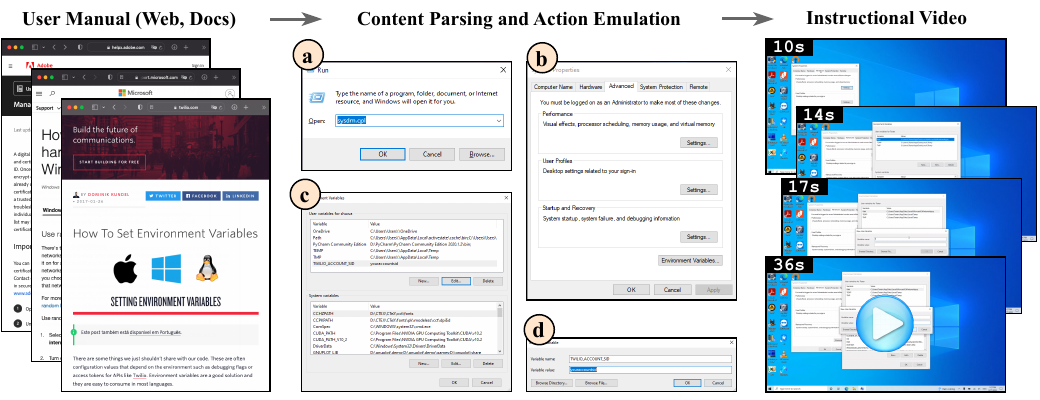}
  \vspace{-0.2in}
  \caption{Given a software/system-related user manual, \TN{} automatically generates a video that presents the manipulation procedure. \TN{} extracts actions (a)-(d) by parsing the text instructions. Then \TN{} leverages the computer vision techniques to emulate these actions. The whole emulation procedure is automatically recorded as an instructional video. \textit{Manual source: Twilio Blog, "How To Set Environment Variables"}.}
  \label{fig:teaser}
\end{teaserfigure}
\settopmatter{printfolios=true}
\maketitle

%%-------------------------------------------------------
\section{Introduction}
\label{sec:intro}

% user manual
A user manual is usually provided to assist users in installing or using a software application, containing both written instructions and associated images for the step-by-step guides. For software vendors, an easy-to-understand user manual can dramatically reduce the cost of technical support~\cite{blackwell1995good}. 
%People nowadays cannot avoid using computers for work or entertainment in their daily lives.
%Many users will consult the user manuals before contacting the IT support team when they encounter new software or software-related problems on their computers. 
% motivation
%Traditional software/system-related user manuals are text-driven documents. After each step, some manuals may provide images to show the software/system screenshots.
%
%why video?
%However, inexperienced users may get confused about the descriptions in user manuals, lead to ambiguity in the instructions.
Compared to the user manuals, instructional videos are more effective and attractive by explicitly guiding users to complete the task and take the next action. Thus, users usually consider the instructional videos as an experience rather than a lesson containing a set of working instructions~\cite{chi2012mixt, tuncer2020pause, zhu2021gif, ramakrishnan2017non, ahmed2012accessible, zhang2017personalized, vtyurina2019verse, zhong2014justspeak}. 
Besides, because the instructional videos provide immediate feedback to the users who just undertake certain actions, users can keep the momentum going during the interaction.
%Because of the advancement of recording devices and online video platforms, an increasing number of user manuals now support videos -- short videos explaining each step and stand-alone videos presenting the entire procedure. Compared with text-driven documents, videos are more effective in showing the exact actions~\cite{chi2012mixt, tuncer2020pause, zhu2021gif}. 

%challenges of generating videos
Nowadays, instructional videos are usually produced by the original software vendors. Besides, enthusiastic third parties may also create some instructional videos to share their experience with the software community containing a large number of users. However, not all user manuals provide video versions for their users due to the extra required effort. Thus, an automatic toolkit is needed to generate these instructional videos automatically for the given software user manuals. 
Existing approaches of video generation~\cite{xu2008animating, willett2018mixed, hayashi2014t2v, kalender2018videolization, chi2020automatic, chi2021automatic, xia2020crosscast} mainly focus on finding the corresponding visual or audio materials for the given descriptive text. However, they cannot generate instructional videos automatically by emulating the actions described in the manuals. 
For example, HelpViz~\cite{zhong2021helpviz} relies on using Android Debug Bridge (ADB) commands on the Android systems to manipulate the GUI elements remotely; however, it cannot be extended to other platforms.
%However, emulating user behaviors on the software and system is relatively convenient. 

% our work
In this paper, we develop a new platform-independent framework called \TN{}, which can automatically convert the software user manuals to instructional videos. As shown in Figure~\ref{fig:teaser}, \TN{} can automatically parse the use manuals to extract the key actions/steps and then generate the instructional videos by simulating these actions.
This framework consists of two main function modules, namely, \textit{action extraction} and \textit{action emulation}. 

% The action extraction module is responsible for extracting a sequence of actions from the user manual by using  natural language processing and computer vision techniques. The action emulation module focuses on operating the software based on the extracted actions and recording the operation procedure in an instructional video.

%
% \noindent {(1)}
The action extraction module is responsible for extracting an action sequence from the user manual by natural language processing (NLP) and computer vision (CV) methods.
% The action extraction module consists of several steps. 
\TN{} first converts software user manuals in various formats to the Markdown format, which contains instructional text and images of GUI interfaces. Then, \TN{} extracts the action list from the manuals by lexical category analysis. According to the extracted actions, \TN{} further utilizes image recognition to locate and extract the corresponding GUI elements from the embedded images. After combining the actions and the corresponding GUI elements, \TN{} can extract the final enriched action list to complete the task.

The action emulation module reproduces the original task in an emulated computing environment, based on the given action sequence from the action extraction module. The extracted action list is translated into the executable code, which can be invoked by the machine directly. To support various platforms, \TN{} uses computer vision techniques, including boundary detection and optical character recognition, to conduct the corresponding actions on a real host to emulate the real user behaviors. Meanwhile, \TN{} generates the instructional video by recording the emulation procedure in the emulated environment. Finally, users can obtain a better understanding of the instructions by watching the generated instructional video.

% evaluation
To evaluate the quality and usability of the instructional videos generated by \TN{}, we conducted a user study with a total of 10 participants (including professionals and non-professionals) and an online survey from 100 respondents. The goal of the user study is to evaluate if our auto-generated instructional videos can assist users in following the software operation procedure. The experimental results show that the total time of completing the software tasks can significantly decrease with the help of the generated instructional videos, compared to that using the original user manuals.
Also, the participants are satisfied with the instructional videos due to the `easy-to-understand' features. The online survey shows that more than 85\% of participants prefer to use our instructional videos rather than the instructional text/images in the user manuals.

%Based on these findings, our framework is capable of effectively generating instructional videos from software user manuals because the users tend to give positive feedback in following the software operation procedure.

% contribution
In summary, we make the following contributions:

\begin{itemize}

\item To our best knowledge, \TN{} is the first system to automatically generate instructional videos from text user manuals, by utilizing both the NLP and CV techniques.
% by extracting the action list, emulating the software operations, and automatically generating instructional videos. 

% \item We implement a prototype of \TN{} based on NLP and CV techniques to generate instructional videos from user manuals. It can be extended to support various types of user manuals. 

\item We perform user studies to evaluate the effectiveness and usability of \TN{}. The experimental results show that \TN{} is promising to generate informative instructional videos. 

\item We find the M2V system can accelerate the cognitive process of non-professionals by converting professional manuals to vivid and easy-to-understand videos.

\end{itemize}

% \noindent {\bf Roadmap.}
% The rest of paper is organized as follows.
% \SW{add the roadmap later, if need.}

\begin{figure*}[ht]
    \centering
    \includegraphics[width=0.92\textwidth]{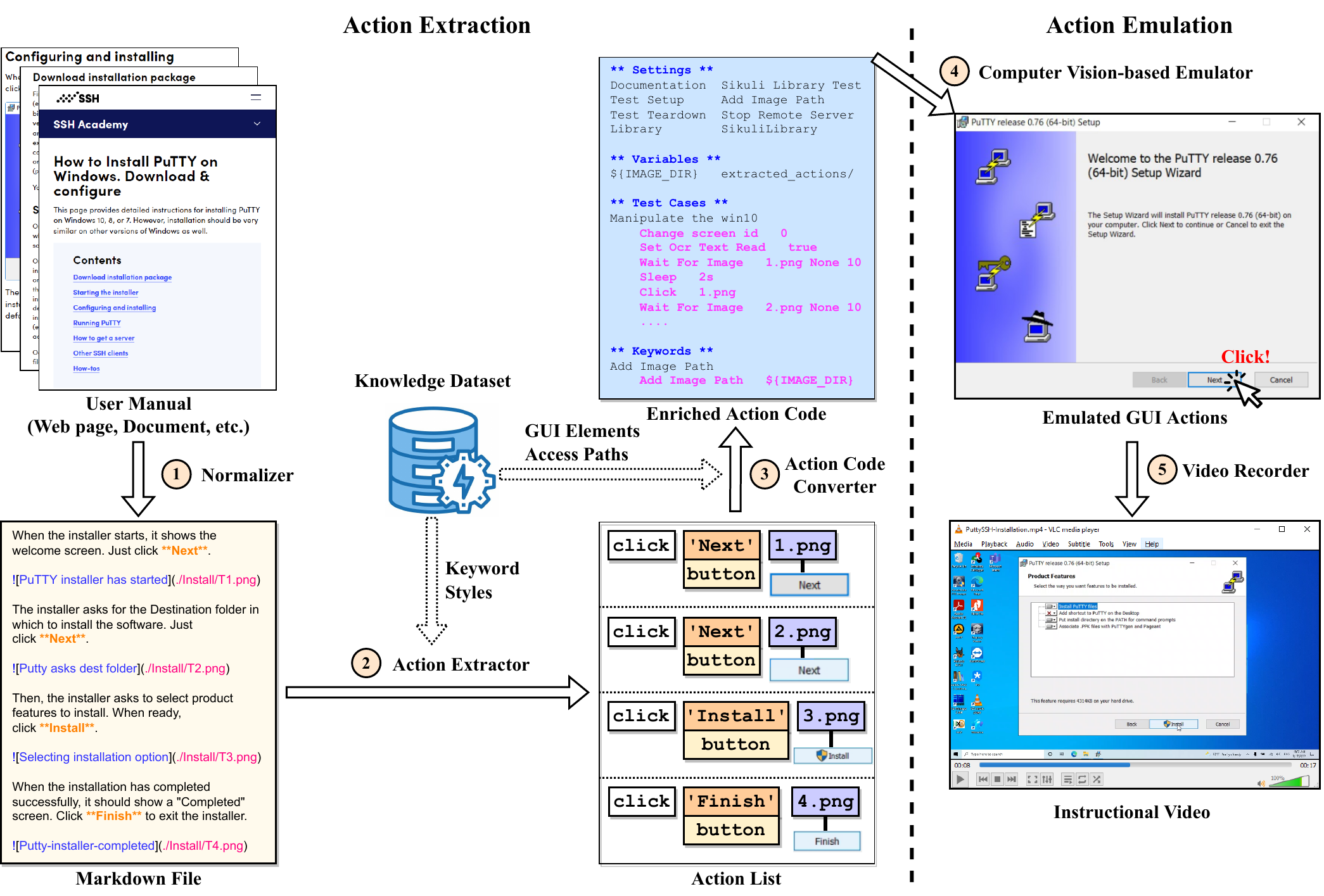}
    % \vspace{-0.05in}
    \caption{The workflow of \TN{}: (1) normalizing  user manuals into Markdown format; (2)  extracting action list from the Markdown-formatted manuals; (3) generating enriched action code from the action list; (4) leveraging CV techniques to conduct the actions of user manuals in an emulation environment; (5) recording the manipulation procedure as an instructional video.} %\textit{Manual source: SSH, ``How to Install PuTTY on Windows".}}
    \label{fig:overview}
   \vspace{-0.05in}
\end{figure*}

% \vspace{-0.05in}
\section{System Design}
\label{sec:design}

%In this section, we present the whole design of the \TN{} framework as well as each module in the \TN{}.

\subsection{Overview of \TN{}}

Figure~\ref{fig:overview} shows the workflow of \TN{}, which consists of two phases, namely, \textit{action extraction} and \textit{action emulation}. 
The action extraction phase converts the software user manuals to the pre-defined action code in three steps.
In {step \ding{192}}, a normalizer converts user manuals in various formats (e.g., web pages, PDF documents) into uniform text, i.e., the lightweight markup language Markdown. In step \ding{193}, an action extractor leverages prior knowledge (e.g., keywords, manual styles) to parse the Markdown files and generate an action list. In step \ding{194}, an action code converter generates an enriched action code that can be directly executed by the action emulator. The grammar of the enriched action code strictly follows the well-defined code structure of Sikuli Robot Framework library~\cite{sikulix}, which includes four parts: settings, variables, test cases, and keywords.
The action emulation phase contains two steps to execute the action code and record the instructional video on a host machine, respectively. In step \ding{195}, a computer vision-based emulator generates real user behaviors by automatically conducting actions based on the action code. In step \ding{196}, a video recorder running on the same machine records the software manipulation procedure into an instructional video.

% \vspace{-0.05in}
\subsection{Action Extraction}

\subsubsection{User Manual Normalizer}

%Before introducing other components of the action extraction module, we need to first discuss the styles of software/system-related user manuals. 
Due to different formats, compositions, and writing styles, it is necessary to normalize the user manuals in advance for further processing.
% User manuals need to be first normalized for further processing because of their different format, composition, and writing. 
% Due to different manual sources and various categories of software, user manual styles differ in their format, composition, and writing. 
User manual files may be presented in multiple formats including web pages (e.g., HTML), documents (e.g., DOCX, EPUB, PDF), and repository readme files (e.g., Markdown), which can be classified into two main categories, namely, \textit{text-only manuals} and \textit{hybrid manuals}. The text-only manuals only include instructional texts and represent GUI elements via descriptive texts. The hybrid manuals contain both instructional texts and illustrated images of the GUI elements.
Moreover, the writing style of user manuals varies for different technical writers. Some manuals may directly indicate the types of GUI elements in their descriptions, while others may use unique symbols to emphasize these GUI elements.
To handle different styles of user manuals, a normalizer is designed to convert different user manual formats into the popular and well-structured Markdown format~\cite{gruber2004daring}, which preserves multimedia (e.g., images and videos) inline with the text. %In this work, images play an important role during the action emulation for identifying the target GUI elements.

% \vspace{-0.02in}
\subsubsection{Action Extractor}
The action extractor is responsible for extracting actions from instructional texts and generating an action list.
% In general, we consider that the user manuals fall into two categories: \textit{text-only manual} that only consists of instructional text, and \textit{hybrid manual} that contains instructional text and corresponding GUI element images.
For the text-only manuals, we obtain a set of action keywords (e.g., ``Click'', ``Select'', ``Type'') and the corresponding target objects (e.g., GUI elements, text).
For the hybrid manuals, we further obtain the images of GUI elements as part of the target objects. We represent each action with a uniform record format as follows.
\begin{center}
{\small 
    \texttt{<Action Keyword>, [Target Type], [Target Object], [Target Path], [Condition]}}
\end{center}
% 可以参考Fig 5: https://storage.googleapis.com/pub-tools-public-publication-data/pdf/c4e1b4c39238bf29f262dee1a43f6b921c09ba7e.pdf
%给上面操作加个例子
\noindent where \texttt{\small Action Keyword} indicates the operations, e.g.,  ``click'', ``right-click'', ``select'', ``type'', and ``input''.
The \texttt{\small Target Type} field indicates the type of an object, which can be used to distinguish the GUI elements (e.g., button, check box, label button). \texttt{\small Target Object} shows the names of the GUI elements or the text content. 
%For the hybrid manuals, we could obtain the images of some GUI elements. 
\texttt{\small Target Path} indicates the stored locations of the images. If an action has a trigger condition, we leverage the \texttt{\small Condition} field to present its trigger event. 
%For each action record, an action keyword must be contained at the head, while other elements are optional. 

Another challenge of \TN{} is to find a uniform rule or a generic pre-trained model to extract all actions from diverse manual writing styles. Some manuals may directly indicate the types of GUI elements in their description. For example, the GUI element of the instruction ``\textit{Click the Button Next}'' is a button with the name of ``\textit{Next}''. Others may use the typographical emphasis to achieve the same function, e.g., an instruction uses the bold font to indicate the GUI element. In the Markdown format, the instruction becomes ``\textit{Click the **Next**}''. In this case, the extractor needs to parse these symbols to locate the target objects.
%; however, the pre-trained model under general corpus cannot parse it correctly. 
%
Besides different text styles, the image styles vary in the hybrid manuals. Some manuals provide screenshots to present the running environments, while others may directly provide the exact GUI element images. Therefore, different rules are demanded to identify the correct GUI elements with various image styles. 
To solve these issues, \TN{} needs to customize action extraction rules for different manual styles. Fortunately, based on the observation that the user manuals from the same source (e.g., author, website) share the same writing style, the action extractor can use a pre-trained deep learning model to process a large number of user manuals with the same styles. %or the user manual styles with a small number of cases, the action extractor leverages the pre-defined rules to parse them. 

% \vspace{-0.02in}
\subsubsection{Action Code Converter}
Given the extracted action list, an action code converter generates an enriched action code with five steps. The generated action code, which contains more details not provided by the user manuals, can assist the action emulator conduct executable actions. 
% The enriched action code contains  more action details that may not be provided by the user manuals. %It consists of five tasks. 
%

%To enrich the action code, the action converter aims to complete the following tasks. 
%\noindent 

\noindent
\emph{{Step 1:} Mapping actions to input peripherals.} In the descriptive instructions, action keywords can infer operations on specific input peripherals (e.g., mouse, keyboard). For example, ``Click'', ``Select'', and ``Check'' are conducted by the mouse click operations to the desired GUI elements. 
% This step can simplify the action types, as well as the action code. 

\noindent 
\emph{{Step 2:} Adding time intervals.} 
The software may quickly switch to another graph interface to conduct the next action. If the interval between two consecutive actions is too short, users need to pause the video to review the previous actions. 
% since users need more time between the two actions to capture the former action from the instructional video,
Thus, we add extra time behind each action for users to better understand each step.  %The length of waiting time intervals can be pre-defined as a constant. %The emulation environment and software response speed may lead to small yet negligible variation on the time interval.

\noindent 
\emph{{Step 3:} Adding access paths to images.} To emulate the user actions, \TN{} identifies the GUI elements (e.g., icons) on the screen via computer vision techniques.
% Software interfaces tend to leverage icons as GUI elements. 
%
However, in the text-only manuals, the extractor can only extract the types and names of GUI elements. Thus, the emulator may not be able to correctly locate the elements via their names. For instance, the Windows menu button is a Windows logo without any text description. When the manual asks users to access the Windows menu button, the  GUI element can hardly be identified by only recognizing the plain text on the screen. 
To solve this issue, we collect common GUI elements in a knowledge dataset and link them to their object names. When actions involve icon-based GUI elements or specific action records, the converter finds the corresponding pre-collected GUI elements and adds their paths into the code.

%For example, the ``Close" button does not contain any text indicator in some software interfaces. The converter translates the action ``Close" to ``click the `Close' button" in the code. The target path of the ``Close" button is derived from the pre-knowledge dataset.

\noindent 
\emph{{Step 4:} Converting abstract actions.} 
 %especially for the system software configuration manuals,
When some actions are well-known to all users, the extracted actions may be incomplete. For instance, the ``Open settings'' action in Windows actually requires the users to perform three detailed actions, namely, searching the keyword ``settings'', locating the ``settings'' item, and clicking the ``settings'' icon. To clarify these actions, the converter replaces those vague abstract actions with pre-defined short operation routines.

%those abstract actions. When the converter encounters those actions, it replaces them with the pre-defined operation routines.

\noindent 
\emph{{Step 5:} Adding action parameters.}
When users are required to fill in customized information (e.g., installation path, IP address), we simply use the default values provided by the software. When the default values are not available, we can customize those values in the emulation environments to maintain the continuity of the emulation process.

% \vspace{-0.05in}
\subsection{Action Emulation}

%\subsubsection{Action Emulator}
The action emulator operates the input peripherals to emulate user activities on the GUI interface based on the action code. Computer vision techniques are leveraged to locate the target objects on the screen, e.g., images and text. For the images, the emulator conducts the template matching to search for the corresponding GUI elements. For the text, the emulator uses the optical character recognition (OCR) and object detection techniques, which can recognize the matching text within the interface and detect the element locations on the screen. By recognizing the operations and the target objects, the emulator executes the actions sequentially.
To avoid developing multiple emulators for different operating systems (e.g., Windows, Linux, MacOS), we select to develop a single emulator on Linux, which manipulates the GUI elements in other OSes via a VNC connection. 
%
%set up the emulation environment as an independent platform (either virtual machine or physical machine) to minimize the effect of emulator on the environment so that the emulator can support various platforms.
%We configure a VNC server on the emulation environment, where the emulator can manipulate the GUI elements via the VNC connection.
%
%\subsubsection{Video Recorder}
Meanwhile, a video recorder running on the emulation environment records the entire operation procedure. The recorder can start and stop according to the time schedule of the action code. Off-the-shelf video recorders are available on different operation system platforms.
% \vspace{-0.05in}
\section{System Implementation}
\label{sec:implement}

%\TN{} is a ge framework to generate instructional videos for various software user manuals, thus it supports different software and systems.

%In the \TN{}, the normalizer, action extractor, action emulator, and video recorder can accommodate the practical requirements of target software, running platforms, manual styles, etc. 

We implement a prototype of \TN{} on a Linux host machine with Ubuntu 20.04 (kernel version 5.4.0). 
Because a large number of user manuals are designed for Windows applications, we use a Windows 10 host as the emulation environment. 
% Therefore, the video recorder runs on the Windows host, 
Despite of the video recorder, all other components of \TN{} are deployed on the Linux host, which connects to the Windows host via a VNC connection. We will release the tool for a public testing.

\begin{figure}[t]
    \centering
    % \vspace{-0.06in}
    \includegraphics[width=0.95\linewidth]{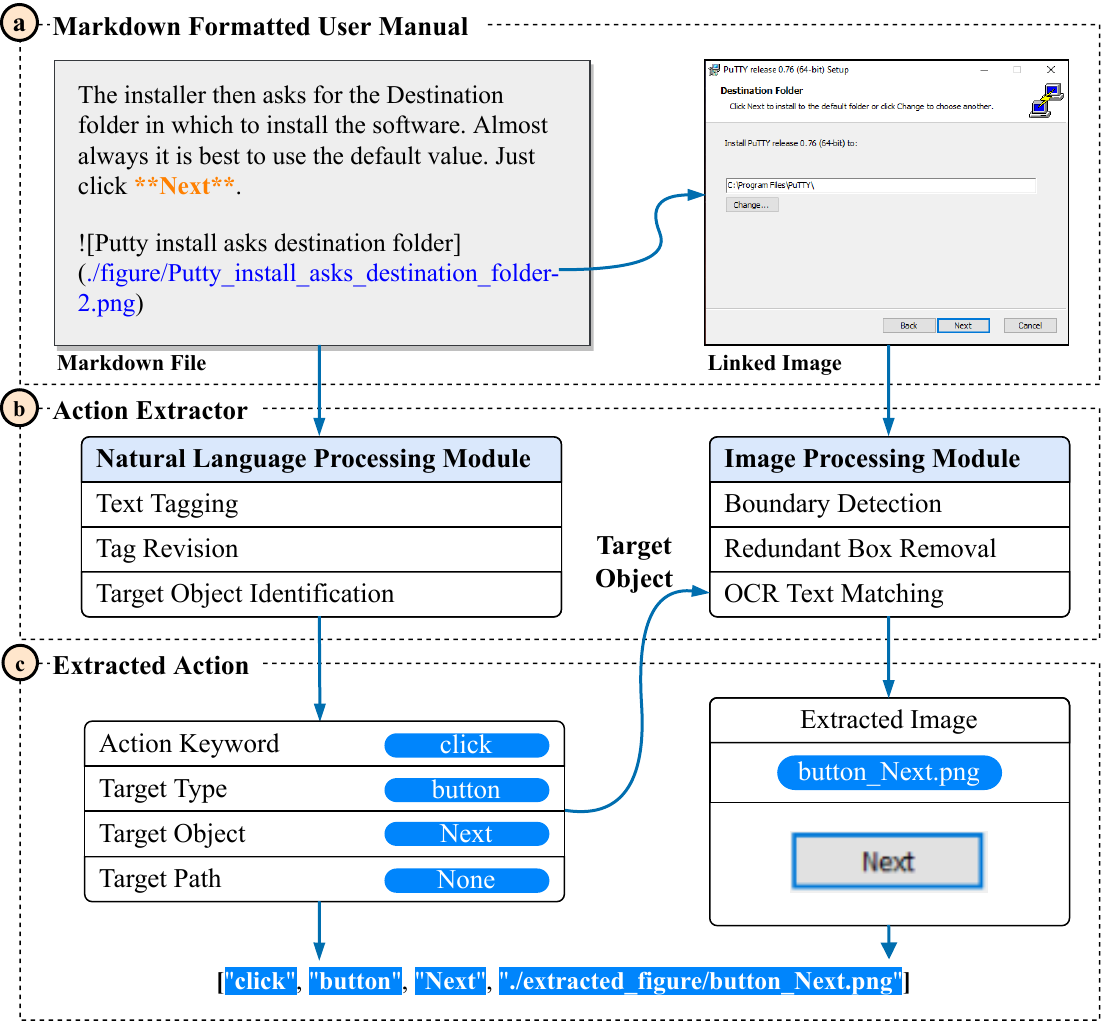}
    % \vspace{-0.2in}
    \caption{An example of automatically extracting actions from a hybrid manual. (a) The action extractor receives the Markdown manual containing text instructions and images. (b) The action extractor processes the text instructions and images respectively. (c) The action extractor combines the extracted information to generate the desired action record.}
    \label{fig:example_action_extract}
    \vspace{-0.1in}
\end{figure}

% \vspace{-0.1in}
\subsection{User Manual Normalizer}
% We use  as 
The normalizer is implemented via \textit{pandoc} 2.5~\cite{pandoc} to convert user manuals in various formats including HTML, DOCX, and EPUB to the uniform Markdown format. For the user manuals in other formats, \TN{} can use  the corresponding Markdown format converters.

% \vspace{-0.1in}
\subsection{Action Extractor}
The action extractor is developed in Python 3.8.10.
The extractor parses the descriptive instructions and obtains the possible GUI elements from the manual images.
% As mentioned above, the Markdown-formatted manuals fall into text-only manuals and hybrid manuals.
% For the text-only manuals, the extractor processes the descriptive instructions to obtain the action list. For the hybrid manuals, the extractor obtains the GUI elements from the manual images and links them to the corresponding actions, besides processing the instructions. 
Figure~\ref{fig:example_action_extract} presents the workflow of the action extraction from a hybrid manual, where we process the text instructions by NLP and handle the embedded images by CV techniques.
% use NLP techniques to process the text instructions and use CV techniques to process the embedded images.

\vspace{0.05in}
% \noindent
{\bf Instruction Processing.}
%\noindent
%{\bf (1) Instruction Processing in Action Extractor:}
%\noindent
%{\bf Text Tagging: }
% The action extractor leverages a pre-defined action keyword list to identify the actions in the instructions.
The basic idea of the action extractor is to identify the pre-defined keywords in the instructions. However, not all matched keywords represent an action. A matched keyword can indicate an action only if it is a verb, e.g., the keyword ``\textit{Input}'' could appear as a noun or a verb in a descriptive phrase while only the verb-form ``\textit{Input}'' can indicate an action. To handle the words with different variations (e.g., ``use'', ``using'', and ``used''), we also apply the stemming methods to convert the words to the root format.
Therefore, we distinguish the types of action words via lexical category analysis. With the \textit{NLTK}~\cite{nltk}, the action extractor splits the instructions into word tokens and then attaches the class tag to each word.

%\noindent
%{\bf Tag Revision:}
% In the practical scenario, the tagging can not reach 100\% accuracy.
In practical scenarios, keyword tagging may not be accurate, especially when involving symbols.
% accuracy is around 70\%. If the manual contains symbols, the accuracy degrades to around 60\%.
The main reason is that user manuals have various writing styles and use professional terms in computer fields, while the \textit{NLTK} is trained with the common corpses. 
% Sometimes the action keywords in the imperative sentences will not be tagged as verbs.
Another reason is that the Markdown format contains symbols, which are the typographical emphasis in the text and can be the indicators of target objects.
% , while they could mislead the tagging results. 
The action keywords ahead of these symbols can be tagged as nouns by mistake. To eliminate these error tags, the extractor needs to re-check the keyword tags and correct the wrong ones.
% automatically recheck the tags of the found keywords and then revise them.

%\noindent
%{\bf Target Object Identification:}
After identifying the action keywords, the action extractor  identifies the corresponding target objects. Typically, a target object should be a noun or a short phrase.
% , following an action keyword. 
When the identified action keywords do not even have the target objects, the action extractor identifies the target objects by conducting a forward search of nouns from the identified action keyword to the next one or to the end of the sentence. Besides, the target objects can be located via the symbols in Markdown, e.g., quotation marks, asterisks. The action extractor can identify these target objects that  usually appear in the descriptive sentences of system configuration manuals. 

% However, our action extractor can also process these situations.
% These situations are also processed by our action extractor. 

\vspace{0.05in}
% \noindent
{\bf Image Processing.}
%\noindent
%\textbf{(2) Image Processing in Action Extractor:}
In the hybrid manuals, the embedded images can provide extra information about GUI elements. Hence, the action extractor processes the images by following the instructions.
% Hence, if an image follows the instruction, the action extractor can request image processing after identifying the action.
There are two types of images in the instructions. One type is the GUI element precisely cropped, e.g., buttons, check boxes, and label buttons~\cite{tuovenen2019mauto}. 
% One image type is the precisely cropped GUI elements, e.g., buttons, check boxes, and label buttons. 
% The audience can learn the exact appearance of the GUI elements mentioned in the instruction.
Because users can learn the exact appearance of the GUI elements, the action emulator matches these GUI elements directly on the screen. Another type is the software interface images that contain the target GUI elements.
% Another image type is the software interface, which contains the target GUI element. 
% These images aim to give the audience an overview of the current status after completing an action. 
In these cases, the action extractor needs to segment the GUI elements from the images.

We leverage the \textit{OpenCV}~\cite{python_opencv} and \textit{Tesseract-OCR}~\cite{python_tesseract} to locate and extract  specific GUI elements from the images.
The action extractor conducts {\em Canny} boundary detection over the images, according to the type of target objects. 
% From the instruction processing, we obtain the type and name of a target object.
% According to the target type, the action extractor conducts the Canny boundary detection in the image.
For example, to detect a specific rectangular button, the action extractor detects all the possible rectangular bounding boxes in the image. 
The boundary detection may provide multiple overlapped detection results due to the low image resolution. 
% The main reason is that the boundary of GUI elements could be obscure, which is caused by the low resolution. 
% Therefore, a GUI element could be detected multiple times, which produces multiple overlapped detected bounding boxes.
To solve this issue, we use Non-Maximum Suppression (NMS)~\cite{neubeck2006efficient} to eliminate the redundant bounding boxes.
Then, the action extractor detects the content in each bounding box via OCR technique and finds out the one with the same name of given target object.
% Then, the action extractor detects the name or content from the remaining bounding boxes via OCR technique.
% Comparing with the target name, the action extractor can find out the correct bounding box for the GUI element.
Similarly, low-resolution images may affect the precision of OCR. Therefore, we calculate the Character Error Rate (CER) between the identified content and the name of the target object and select the bounding box with the highest CER score. 
% We calculate the Character Error Rate (CER) of the identified text with the target name. 
% As we know, the target GUI element is in the processed image.
% Therefore, the bounding box with the highest CER score has the largest probability to be the target GUI elements.
With the correct bounding boxes, the action extractor can corp the images of GUI elements and add these image paths into the action record.

%However, some icon-based GUI elements without description text may show aside on the screen. The OCR technique cannot detect them from the provided images, thus the action extractor cannot extract them. Currently, we rely on the pre-knowledge data to fill this gap in the action converter.

\subsection{Action Code Converter}
% Similar to the action extractor, the action converter is also developed in Python.
The action converter transforms the action list into an enriched action code, i.e., the Robot Framework code, which has easy-to-use tabular test data syntax and utilizes the keyword-driven testing approach.
%The Robot Framework code ensures the action converter simple and effective.
The converter translates the action records into the test cases in sequence, takes action keywords as the code keywords, and transforms the target object information as the parameters.

We enrich the generated code by introducing the pre-knowledge data, which includes the waiting time slots, the images of icon-based GUI elements, the groups of action sequences, and the action parameters. The action converter inserts the waiting time slots (i.e., pauses) between two consecutive actions. For common actions, we set the general waiting time to 2 seconds, considering the software response speed and action complexity~\cite{gomez2013reran}.
% For common actions, the time slot could be short hence we can set it according to the software response speed and the complexity of these actions. 
% In our test environment, we set the general waiting time slot to 2 seconds. 
More waiting time may be required to complete some complex actions, e.g., the installation time varies for different applications.
% Some actions require more time to complete a task, 
For these actions, we obtain the target object of the next action as the completion indicator. In the final code, the converter configures the waiting time so that the next action will not be conducted until the target object appears.
% the waiting time of the current action ends 
% if the target object of the next action appears.
% Here, we mainly discuss how to set waiting time slots, while we collect other components according to the features of the corresponding software.

\begin{figure}[t]
    \centering
    \includegraphics[width=0.85\linewidth]{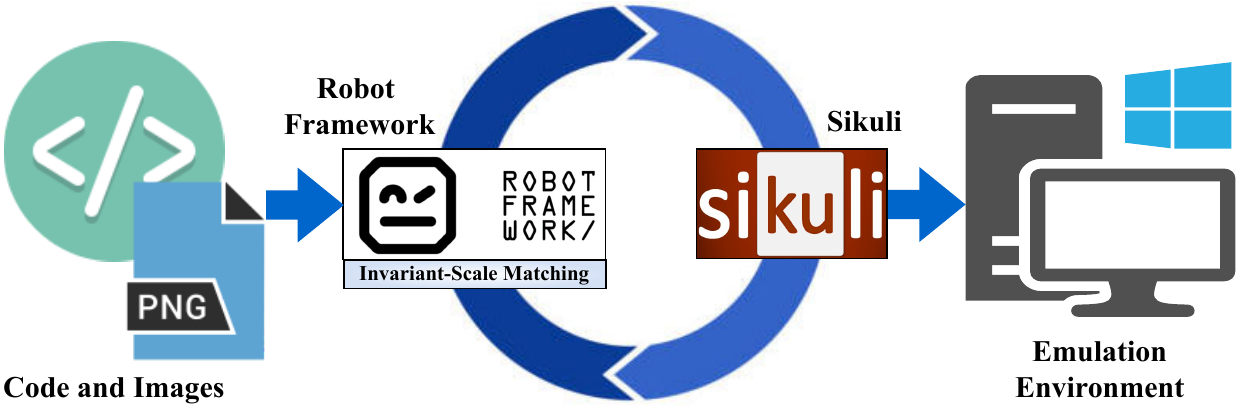}
    \vspace{-0.1in}
    % \caption{The implement of action emulator. Our action emulator is based on Robot Framework and Sikuli. The Robot Framework parses the input action code and invoke the Sikuli model to execute the desired actions in the emulation environment. We also add an invariant-scale matching module to Robot Framework to support image matching in different scales.}
    \caption{The implementation of action emulator.}
    \label{fig:emulator-imp}
    \vspace{-0.15in}
\end{figure}

%\subsection{Action Emulation Module}

\subsection{Action Emulator} 
% 可以写个算法
In Figure~\ref{fig:emulator-imp}, the action emulator consists of the \textit{Robot Framework} 4.1~\cite{robotframework} and \textit{Sikuli} 2.0.5~\cite{sikulix}. Robot Framework is a Python-based, extensible, and keyword-driven automation framework, which can be used for test automation and robotic process automation. Robot Framework has a rich ecosystem consisting of various generic libraries and custom libraries. We use the Sikuli library of Robot Framework to manipulate the GUI elements on the screen. Sikuli uses image recognition algorithms powered by OpenCV to identify GUI elements and uses OCR to identify text.

A restriction of Sikuli is that the candidate image must be the same width and height (pixels) as the target image~\cite{yeh2009sikuli, chang2010gui}. When the images in manuals do not match the screen resolution, it may lead to the failure of image matching.
% It is difficult to ensure the images share the same resolutions with the screen, which could easily cause failure in image matching.
To solve this issue, we develop an invariant-scale matching module for Robot Framework working with the Sikuli module. The basic idea is to loop over the input image at multiple scales to find a matching one.

In Algorithm~\ref{alg:action_emulation}, the Robot Framework receives the code and related image.
If the input image has the same resolution as the screen, the Robot Framework forwards the image to the Sikuli module, which emulates the corresponding action on the identified screen area (Line 4).
If the input image has a different resolution from the screen, the Robot Framework drives the Sikuli module to capture a screenshot from the emulation environment. Then, the screenshot and the input image are sent to the invariant-scale matching module (Line 6), which searches the matching area from the screenshot under various scales (Line 13) and returns the corresponding coordinates to the Sikuli module via Robot Framework. With the coordinates, the Sikuli module can continually emulate the actions on the correct GUI elements (Line 8).

\begin{algorithm}[htb] 
\caption{Action Emulation with Input Image} 
\label{alg:action_emulation}
\SetKwProg{Fn}{Function}{:}{}
\SetKwFunction{FEmulator}{Robot\_Emulator}
\SetKwFunction{FMatch}{Invariant\_Scale\_Match}
\small
\LinesNumbered
\SetNoFillComment
\KwIn{~~\\
~~~~ $A$: Action to be executed on the GUI element\;
~~~~ $I$: Input image for searching the corresponding GUI element\;}
% \KwOut{Execute the Action with Sikuli;}

\Fn{\FEmulator{$A$, $I$}}{
$S$ = Sikuli.screenshot();

\If{$I$.resolution() == $S$.resolution()}{ 
    \tcc{Same Resolution}
    Sikuli.execute($A$, $I$);
} \Else { 
    \tcc{Different Resolutions}
    $C$ = Invariant\_Scale\_Match($I$, $S$);
    
    \If{$C$ is not \texttt{None}}{
    Sikuli.execute($A$, $C$);
    }
}

}

\vspace{1mm}

\Fn{\FMatch{$I$, $S$}}{

    $found$ = \texttt{None};
    
    $similarity$ = \texttt{None}; 
    
    $coordinate$ = \texttt{None};

    \tcc{Loop Over the Scales}
    \For {$scale$ in range($Min$, $Max$, $Step$)} {
    
    \tcc{Resize the Input Image}
    $I_r$ = resize($I$, $scale$);
    
    \tcc{Conduct the Template Matching}    
    ($similarity$, $coordinate$) = matchTemplate($I_r$, $S$);
    
    \tcc{Update the Matched Area}
    \If{$found$ is \texttt{None} or $similarity$ $>$ $found$[0]}{
        $found$ = ($similarity$, $coordinate$);
        
    }
    
    }
    \KwRet $coordinate$
}
\vspace{0mm}
\end{algorithm}
% \vspace{-0.1in}

\subsection{Video Recorder}
We leverage a screen recorder \textit{RecForth}~\cite{recforth} in a Windows 10  based emulation environment. \TN{} sends the start hotkey (\texttt{F2}) to start the screen recording.
When the emulation is completed, \TN{} sends the stop hotkey (\texttt{F1}) to stop the screen recording.
These two actions can also be conducted by the emulator by adding the recording start/end to the action code.
\section{User Evaluation}
\label{sec:ueval}

To evaluate the videos generated by \TN{}, we conduct two user studies to understand how well the videos can improve users' comprehension. The first study investigates how instructional videos help users complete tasks in practical scenarios. The second study investigates the audiences' subjective feelings after watching the automated-generated instructional videos.

\subsection{Study I: In-person Experiments}

\subsubsection{Study design}
In the first study, we investigate if our generated videos can effectively assist users to follow the required software tasks. Because \TN{} mainly focuses on the video generation of software installation and manipulation, we conduct a within-subject interview study with 10 participants from various computing backgrounds (i.e., 4 majored in computer science while 6 majored in non-CS fields). We will consider the CS majored participants have the similar understandings with the professional software developers.
All participants have basic understanding and experience in configuring software, but they did not complete the test tasks before. 
All participants are above the age of 21.
Because we neither record age or ethnicity information nor study any human behaviors, our study does not involve the ethical issue or apply to IRB.

\vspace{0.05in}
\textbf{Materials.}
We select six user manuals from two main categories: software manipulation and system configuration. 
The software manipulation includes OpenVPN installation (\textbf{M1}), LibreOffice Marco editing (\textbf{M2}), and Adobe Reader trust configuration (\textbf{M3}). These software-related user manuals are downloaded from their corresponding official documents. The system configuration manuals record Windows 10 configuration methods, including system environment setting (\textbf{M4}), IP address setting (\textbf{M5}), and network discovery configuration (\textbf{M6}). These system-related user manuals are downloaded from the knowledge platform - Dummies (\texttt{\small https://www.dummies.com/}),
% \footnote{https://www.dummies.com/} 
and the technique blog - Twilio Blog (\texttt{\small https://www.twilio.com/blog}).
% \footnote{https://www.twilio.com/blog} 
Different sources can provide different styles of user manuals. 
The system-related manuals from Dummies contain fewer GUI element images (one or two images at the end of manuals to show the final status) than other software-related user manuals.
In the experiments, \TN{} treats these system-related manuals as text-only manuals and treats other manuals as hybrid manuals.
Therefore, although the case study only include six software tasks, the video generation from these six user manuals can test all techniques used in \TN{}.
We believe \TN{} can be generalized to other different user manuals if it has a good performance over these six typical cases.

These user manuals not only have different styles but also provide different difficulties for the participants. 
Within three software manipulation tasks, the actions of {M1} are straightforward since the participants only need to click several buttons. {M2} and {M3} increase the difficulty since the participants have to switch between multiple windows to complete the task. {M2} also requires participants to type in a small piece of code provided by user manuals, which is new to the participants with limited programming knowledge. We consider different task styles because the real tasks can have various difficulties when we apply \TN{} on general tasks.

For three system-related user manuals, they are basic tasks in the Windows systems; however, normal users may not usually encounter the tasks of {M4} and {M6} (their difficulties are between {M1} and {M2}) in their daily life.
Considering that users may set manual IP addresses for their networks, we regard task {M5} as an easier task. 

For each user manual, we successfully create an instructional video (around 1 minute) using \TN{}. We intend to limit the length of each generated video for this experiment since fewer viewers watch the entire length of longer videos and their engagement tends to drop after 2 minutes~\cite{fishman2016long}.
When participants lose their attention to the video, it is difficult to tell if the generated video contains clear information according to their feedback. Thus, we pick up these user manuals with an average of 10 steps.

\vspace{0.05in}
\textbf{Procedure.}
In this experiment, each participant followed all six user manuals to complete the corresponding tasks.
To ensure the consistency of the experiment, we interviewed the participants one by one and provided a Windows 10 laptop (CPU: Intel Core i7-7700; memory: 32 GB) as the test environment.
The required software of Adobe Reader and LibreOffice has been pre-installed on the test platform. All the participants completed the tasks on the same laptop.

We counterbalanced the videos and text instructions so that each participant can access three videos and three groups of text instructions. For example, a participant watches videos of {M1}, {M3}, and {M5}, and then she needs to read text instructions of {M2}, {M4}, and {M6}. 
We tried to minimize the bias by evenly distributing tasks between CS and non-CS majored participants. Meanwhile, we also balance the task difficulties in each group.
Note that each video/text combination is assigned for a CS or non-CS participants evenly since we tend to minimize the impact from participants' backgrounds (e.g., majors, previous experiences). Since participants may not be familiar with all these tasks, we allow the participants to read text instructions or watch videos more times. After watching a video or reading a group of text instructions, the participants followed it to re-do the task in the test environment.

We measured the {\em video watching time}, {\em instruction reading time}, and {\em task manipulation time} for each participant. We informed the participants that switching between the instructions and the test environment was allowed for completing a task because participants may not remember all the steps.  We also measured the {\em switching frequency} for each task. The participants were allowed to take a break after each task. After completing each task, each participant was asked to provide verbal comments about the text instruction or the generated instructional video. 
The total experiment time for each participant lasted about 50 minutes.

\begin{figure*}[t]
    \centering
    \subfloat[Participants' time usage for completing each task.]
    {\includegraphics[width=0.45\linewidth]{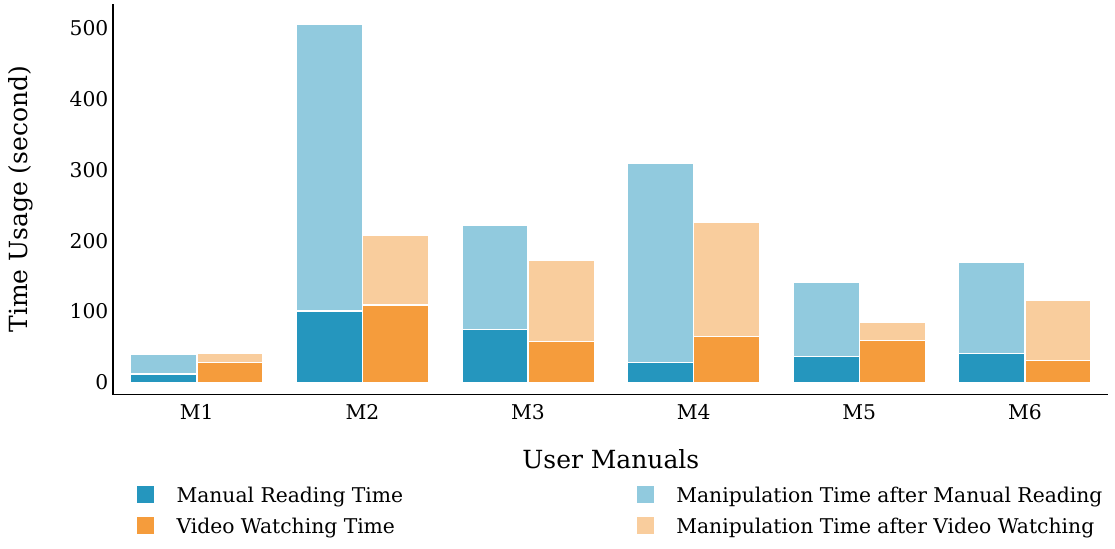}}
    \hspace{0.15in}
    \subfloat[Switching frequency between software and instructions.]
    {\includegraphics[width=0.45\linewidth]{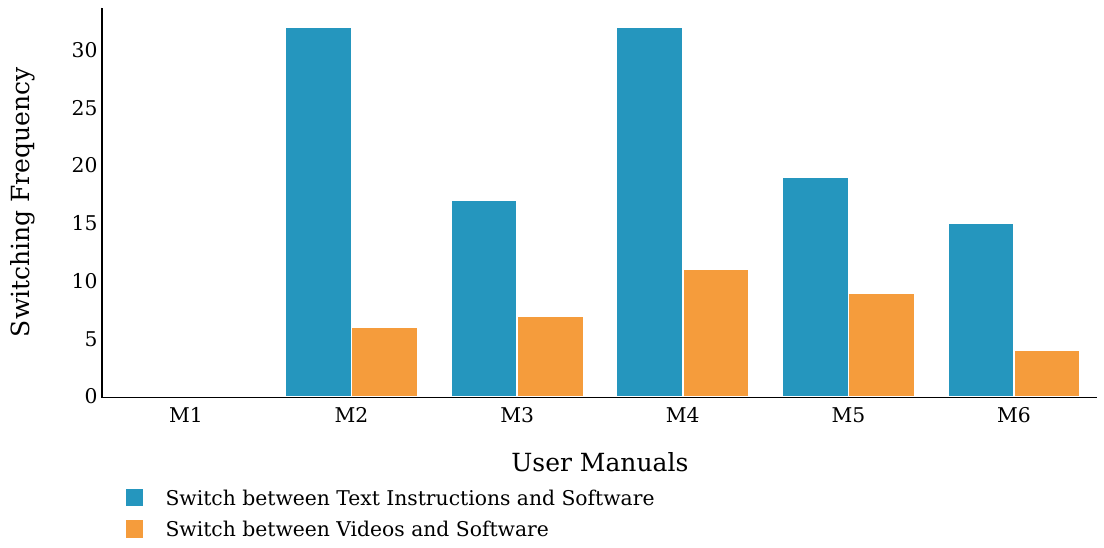}}
    \caption{The experimental results of in-person experiments.}
    \label{fig:time-cost}
    \label{fig:switch-time}
    \vspace{-0.1in}
\end{figure*}

\subsubsection{Results}
All the participants walked through three user manuals in text and other three user manuals in videos. After completing the corresponding tasks, they provided their feedback. Therefore, we also consider both subjective and objective metrics to verify the effectiveness of \TN{} in improving the understanding of learners.

\vspace{0.03in}
{\bf Task completion time.}
First, we analyze the spending time in hands-on operations for each task.
Figure~\ref{fig:time-cost}(a) presents the average time used to complete each task under text instructions and instructional videos.
In general, text instructions are efficient for glancing through a procedure. 
The average reading time on text manuals is 9.5 seconds, less than the average video watching time.
That means the participants read the text instructions faster than watching the whole video, especially for the simple instructions, such as {M1} (software installation).
Note that the watching time on an instructional video is directly related to the video length.
Based on the observation above, it is necessary to reduce the video length to improve the watching efficiency when generating videos with \TN{}.
However, too short videos may lose key information and make it hard for users to fully understand.
Therefore, we understand that deciding video length is a trade-off between user comprehension and watching efficiency.
If the writing of text instructions is not straightforward, the participants may need to read the text more times, which prolongs the reading time. 
Therefore, the reading time of text instructions is longer than the video-watching time in {M3} and {M6}, which involve too many interpretive sentences.

During the hands-on manipulation, the participants can spend less time completing all six tasks when following the instructional videos.
The average manipulation time after video watching ($\sim$100 seconds) is less than that after manual reading, proving that the generated instructional videos can greatly improve user operation efficiency.
Even considering the time spent on instruction acquisition (i.e., manual reading or video watching), the total task completion time with instructional videos ($\sim$90 seconds) is less than that with manual instructions.
From the experimental results, the participants watching videos can save more time on task completion, which shows the effectiveness of \TN{}.
The time difference is even remarkable if the task is relatively complicated.
For example, {M2} (LibreOffice Marco editing) involves multiple complicated operations, including menu selection, item creation, and code insertion. 
Although the text manual is enough detailed, users will still spend a lot of time finding the matching button positions, making various attempts, and determining code insertion location. 
However, the instructional videos can present more clear operations and thus save the users' time in finding and trying.

\vspace{0.05in}
{\bf Switching frequency.}
We understand that participants may not remember all the steps by viewing the instructions only once and may look back at instructions during manipulation.
We regard a review of the instructions as a ``switch'' during the operations and record the switching frequency for each task. 
Figure~\ref{fig:switch-time}(b) shows the average number of switching to videos is 13, less than that of switching to text.
That is because the instructional videos can inform users of the operations visually.
The vivid video form can make users remember almost all the steps and reproduce the operations in a short time.
Users only need to follow the steps in the videos to complete the tasks, instead of understanding the text manuals and thinking about how to perform the next step.

For a simple task such as software installation ({M1}), there is no need for participants to switch from software to the instructions. 
The participants can easily remember to click several buttons to complete the task.
Also, the software already provides sufficient prompts on the user interface and participants may already have enough prior knowledge for these daily operations.
For the other five tasks, the participants' switching frequency toward text instructions is higher than that toward videos. 
The main reason is that participants do not remember the unfamiliar GUI elements, thus they need to repeatedly check them and spend more time finding them on the screen. 
With the instructional videos, they can just remember the locations of the GUI elements used in the tasks, even though they still do not remember the exact names of these elements. 
The experiments on switching frequency show that instructional videos can increase the locating success rate and decrease the operation time caused by switching to manuals.

% \begin{figure}[t]
%     \centering
%     \includegraphics[width=0.58\linewidth]{figures/switch_time.pdf}
%     % \caption{The participants' switching frequency between software and the corresponding manuals for completing each task. Obviously, instructional videos make it easier for users to remember more details of operations.}
%     \vspace{-0.1in}
%     \caption{The participants' switching frequency between software and the corresponding manuals for completing each task.}
%     \label{fig:switch-time}
%     \vspace{-0.2in}
% \end{figure}

\vspace{0.05in}
{\bf Comments from participants.}
During the user study, participants were asked to explain their feelings about the text instructions and instructional videos after finishing all the tasks.
We gathered all the opinions and summed them up into a few common comments.
We find that users are more satisfied with our generated instructional videos compared with the original manual text.
For the opinions on text instructions, we sum up the following common comments.

% 3th is ok, but not care about the detailed info
% Do not know where is Windows Run prompt

% need to highlight the button and optional

% When view videos, not need to care about the selected strings, provide location and icon, speed up the process

{\bf C1. }
``\textit{I cannot remember the detailed steps, especially when the sentences are too long.}''

{\bf C2. }
``\textit{The manual text explains too much `why', but I just want to know `how' to do it.}''

{\bf C3. }
``\textit{It is not easy to find the corresponding options when I face a button list or attribute list in preferences.}''

{\bf C4. }
``\textit{It is better to highlight the buttons in text or provide the button icons so that I can quickly find them.}''

{\bf C5. }
``\textit{I don't know how to launch the `Command Prompt' in Windows.}''

From {C1} and {C2}, we know that users only want to know how to perform the operations and complete the tasks in the simplest manner.
Although the explanations may be necessary in some cases, the lengthy sentences usually bore users and decrease the operation efficiency.
From {C3} and {C4}, we find that the manual text may not provide sufficient support for complex operations, e.g., some users are hard to find the correct icons in a complex user interface.
Sometimes, users only need to know ``where'' to click the icon instead of the description of the icons. 
These issues can be easily solved by using instructional videos.
From {C5}, we also find text instructions may not be friendly to the users in other fields since some basic steps of  professional operations may be omitted.

For the generated instructional videos, the participants made the following comments. 

{\bf C6. }
``\textit{When viewing videos, I do not need to care about the selected options; I just need to click the same icons and insert targeted text in the same locations, which speeds up the whole process.}''

{\bf C7. }
``\textit{The videos could be better if there are instructions or narration presented in the videos.}''

From the comments above, we find it is easier for users to follow the instructions via videos since videos can provide a set of exact actions.
The specific locations of these actions provided by videos will greatly improve the user experience.
In addition, video and text can be combined to further improve the user's understanding of specific operations.

\subsection{Study II: Audience Survey}

\subsubsection{Study design}
In the second user study, we collect end users' subjective and direct feelings about text manuals and instructional videos through a large-scale questionnaire.
Our goals were to understand (1) whether the audiences would accept watching the generated instructional videos and (2) whether the automated-generated instructional videos can assist the audiences to understand the content better.
To understand how the general audiences perceive our generated videos, we conduct an online survey via the crowdsourcing platform - Amazon Mechanical Turk.\footnote{https://www.mturk.com/} 
After posting our survey, we received 100 unique responses from anonymous users.
We check the response patterns manually to confirm there is no malicious filling on the questionnaire.

\vspace{0.05in}
\textbf{Materials.}
For this online survey, we keep using the same text manuals and generated videos in Study I to provide consistent quality for analysis.
Because this user study is a comparative experiment, we divide all materials into 6 groups.
Each comparative group contains a text manual and a corresponding generated video so that users can directly perceive the similarity and differences between these two forms of user manuals.

\begin{figure*}[t]
    \centering
    \includegraphics[width=0.98\textwidth]{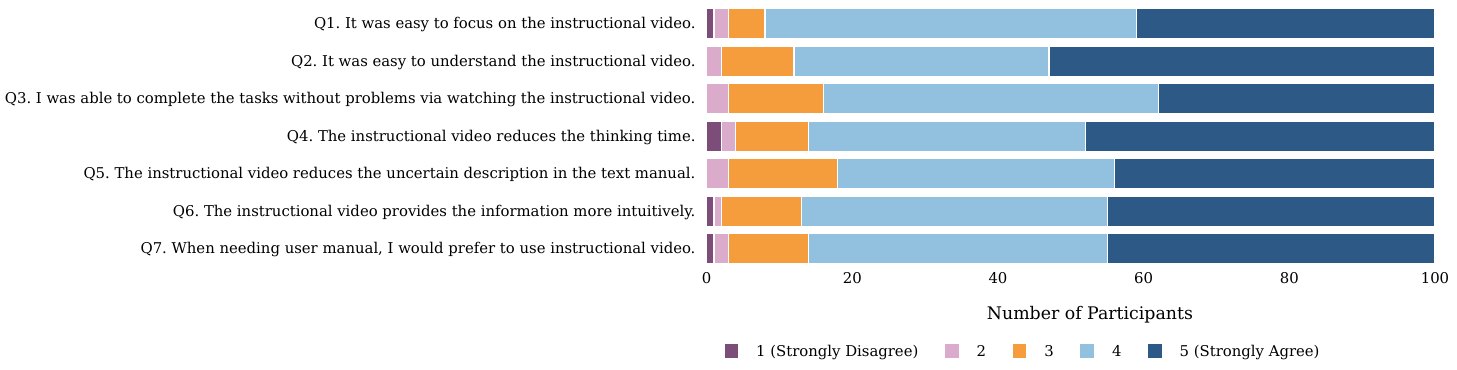}
    \vspace{-0.1in}
    \caption{Participants' responses to seven video-quality questions at 5-point Likert scale in our Study II. Each participant answered the questions after viewing six user manuals and the videos generated by \TN{}.}
    \label{fig:survey}
    \vspace{-0.1in}
\end{figure*}

\vspace{0.05in}
\textbf{Procedure.}
This experiment consisted of two parts. 
First, the participants view all videos and text instructions in these six comparative groups of user manuals.
Then, we asked them to fill in a post-study survey about their experiences with these six manuals in both forms, and we collected demographic information. 
Specifically, we asked them to answer seven 5-point Likert-scale questions as follows.

\vspace{0.03in}

\textbf{Q1}. It was easy to focus on the instructional video. (Easy to focus)

\textbf{Q2}. It was easy to understand the instructional video. (Cognition)

\textbf{Q3}. I was able to complete the tasks without problems by watching the instructional video. (Correctness)

\textbf{Q4}. The instructional video reduces thinking time. (Efficiency)

\textbf{Q5}. The instructional video reduces the uncertain description in the text manual. (Details)

\textbf{Q6}. The instructional video provides the information more intuitively. (Brief)

\textbf{Q7}. When needing a user manual, I would prefer to use an instructional video. (Preference)

\vspace{0.03in}

The scale is ranged from Strongly Disagree (1) to Strongly Agree (5).
Their short names are in parentheses. These 7 questions are designed from different dimensions. {Q1} indicates if the video is attractive to users and {Q2} indicates the cognitive effectiveness of instructional videos. Besides the cognitive feedback (Q1, Q2), we also obtain the feedback on the potential task completion, including the correctness of task completion (Q3), the efficiency to finish task (Q4), the included details (Q5), and video brief (Q6). Finally, we also collect the preference of users (Q7).
The participants need to watch videos, read instructions and answer these seven questions.  Hence, we set the maximum completion time of the questionnaire as 25 minutes. 

\vspace{-0.05in}
\subsubsection{Results}
Figure~\ref{fig:survey} shows participants (general audiences) agreed that our generated videos were easy to follow.
Most of the respondents confirmed that the video form is easier to focus ({Q1}) and easier to understand ({Q2}).
The participants in Study I also gave out similar comments on the instructional videos ({C6}), since users can just follow the steps in the videos and take no effort to understand.

The respondents thought they can complete the task correctly ({Q3}) and effectively ({Q4}) with the videos, since the video is a more direct form to present all the needed operations to users. The exact actions provided by videos can reduce the ambiguity brought by the textual descriptions and guarantee the correctness of operations. Therefore, users can spend less time completing the same tasks by viewing the instructional videos. 
In addition, according to Study I, participants can complete complicated tasks faster with lower switching frequency under the guidance of videos. The respondents agreed the information provided by videos contains more details ({Q5}) and is briefer ({Q6}).
From Study I, we know that more details are present mainly in the shown positions of options. It is consistent with the comments {C1} and {C2} that videos are briefer than text manuals. Overall, they prefer to use instructional videos as the guide if possible ({Q7}).

For most questions, the participants choosing ``Strongly Agree'' (accounting for about 40\%) are more than those choosing ``Agree''. 
We noticed that fewer respondents strongly agreed with the {Q1} and {Q3}.  That is because the current generated videos do not contain any instructions or text titles. Some respondents expected to read some short instructions or hear narration to explain the procedure presented in the videos. %We discuss this issue in Section~\ref{sec:discuss}.

\section{Threats to Validity}

\subsection{Internal Validity}

We analyze the independent variables that could affect the internal validity of \TN{} tool.
First, the manual length will not affect the performance of \TN{} because the action code is in sequential order and is not limited by the number of operation steps. Second, the \TN{} will not be limited by the operation units because software operations are composed of some basic operations (e.g., click, hold, drag) that can be fully emulated by Sikuli action code. Third, the manual types will not affect the effectiveness of \TN{}. In the design, we consider two types of text manuals, i.e., pure text manual and hybrid manual (text with figures). Also, in the evaluation, we verify that \TN{} can work on both software manipulation manuals and system configuration manuals.

\subsection{External Validity}

To generalize the \TN{} to other manuals, there are two main threats to its validity.
First, some manual is only open to the professional users thus prior knowledge is needed to parse the manual.
For example, some items have specific explanations on professional fields (e.g., ``select channels'' in Adobe Photoshop) or some software operations are well-known in the specific fields but rarely used in the common software manipulation (e.g., ``equalization'' in audio processing software).
Although, \TN{} has a database that can include these specific operations, it cannot cover all the possible cases in the real world. Therefore, a beginner-friendly manual writing style is preferred in the input text manuals.
Second, some manuals, especially system configuration manuals, can provide a set of different paths. For example, after selecting different configurations, the manual will go to different steps.
However, since we can convert the manuals into enriched action code, this property is similar to the ``if-then-else'' branch in source code. We can split the action code into multiple basic operation block and record videos separately. After generating an interactive video, user can watch the related subsequent videos by selecting their preferences. We will leave this to our future work.

\section{Discussion}
\label{sec:discuss}
% Overall, we received positive feedback on our generated videos.
% Participants can successfully understand and follow the videos to complete the corresponding tasks.
% The general audience found the videos easy to watch and understand.
% Below we describe the limitations and the opportunities of our approach.

\textbf{Visual Instructions and Narration.}
%Our current solution creates the instructional video by recording the software/system manipulation procedure in the emulation environment.
The generated video by \TN{} does not contain visual instructions or narration.
According to the feedback from our experiments, adding visual instructions and narration to the videos could be more convenient for users to link the specific actions in the videos to the corresponding instructions in the user manuals. In future work, we plan to add these features to our videos by integrating HowToCut~\cite{chi2021automatic}, which can automatically add visual instructions and a voice-over to the videos according to the instructions.
%We will tune the technique for the software/system-related user manuals and integrate it into \TN{}.

%\textbf{Video Length.}
%As pointed out by Fishman~\cite{fishman2016long}, fewer viewers watch the entire length of longer videos. The generated video should be as short as possible. In practical scenarios, the length of a user manual and the number of manipulation steps increase with the complexity of the task. For \TN{}, the length of the generated video also increases at the same time. Besides the complexity of the task, the execution time for each step also influences the video length. For some software, the execution time of one step could be relatively long and prolong the length of the generated video. For example, when installing the software, \TN{} emulator would press the button "Install" and wait for its completion to start the next action. The installation time of small-size software is short, while the time of large-size software could be quite long (more than 10 minutes). The overlong waiting time for a simple action is meaningless for the audience. One solution is to pause the recording during the installation and resume the recording in the next stage. It may skip the status switch procedure. To keep the switching procedure, it is necessary to remove the redundant waiting time by shrinking the generated video.  In future work, we aim to research automatic content-aware video length shortening.

% \vspace{0.05in}
\noindent \textbf{Automatic Parser Selection.}
There are some pre-trained parsing models~\cite{li2020mapping, she2014teaching, branavan2009reinforcement, manuvinakurike2018edit} that can parse the natural language instructions
into executable actions. 
Their design goal is to parse several specific types of user manuals. 
Because \TN{} supports various types of software/system-related user manuals, the action extractor has to leverage different text parsers to process instructions and extract actions. 
Currently, we manually select the appropriate parsing models or create parsing rules for user manuals from different sources.
We plan to make \TN{} recognize the styles of input user manuals and select the appropriate parsers automatically in the future.

% \vspace{0.05in}
\noindent \textbf{GUI elements detection.}
Our action extractor uses the computer vision-based method to detect GUI elements. It leverages the Canny boundary detection to acquire primitive shapes and regions of GUI elements and then uses OCR to detect the GUI elements. Currently, there are some deep learning-based solutions archiving success in object detection research areas, such as Yolo~\cite{bochkovskiy2020yolov4}, UIED~\cite{xie2020uied, chen2020object}. 
To use these deep learning-based solutions in our action extractor, a large data set of GUI elements from various desktop software is required for the object detection model training. We leave the creation of a data set of software/system GUI elements as our future work.

% \textbf{Long-term Memory.}
% As the experimental results present, our generated video can decrease the overall hands-on time and switching frequency of users.
% In a longer time span, the easy-to-remember feature of videos may also bring accumulated benefits. Since the video is easy to understand and remember, users do not need to review the video or only part of it when using the software again. So the overall time for using this software multiple times in the future can be significantly less. But for a traditional user manual, people may need to refer to it every time. This issue requires long-term study in the future.

% \vspace{0.05in}
\noindent \textbf{Version Difference.}
Our emulation environment deploys the same version of software and OS as mentioned in the user manual. However, in real scenarios, users may use different versions of software or OS, and the version difference can cause big changes in the GUI interface.  For example, the GUI interface of Windows 10 is quite different from Windows 8. If \TN{} needs to generate videos based on the user-desired version, it may encounter failures in the matching of the GUI element. We choose to omit the images and only use the text instructions with some pre-collected images to walk through them. If the software manipulation logic is unchanged, \TN{} still can extract the correct action sequence. Thus, we can generate a Windows-based instructional video of LibreOffice from its Mac OS-based user manual.

% \vspace{0.05in}
\noindent \textbf{Usability.}
\TN{} can be integrated into the manual websites to automatically convert existing text manuals into instructional videos.
The instructional videos can be greatly beneficial to those in need of help with a manual, especially any persons with disabilities that are more visual learners.
Also, the techniques used in \TN{} can generate real user behaviors, which have wide security-related applications in enhancing machine fidelity, e.g., enhancing the fidelity of honeypots~\cite{songsong2023honeypot} 
or malware sandboxes~\cite{liu2022enhancing}. 
In our current work, parsing manual relies on the named entity recognition; however, with a well-trained large language models (LLM), it is possible for \TN{} to first summarize the user manuals for better action extraction and make this process more intelligent.

\noindent \textbf{Insights into Manual Writing.}
Our work can also provide insights into the manual writing.
In our paper, we point out the biggest challenge is that manual writers usually assume the readers have some background knowledge. For example, “Open the Settings panel” would be harder for the machine to understand because the model needs to know where can it find the entry of the panel. Instead, the writers should explain clearly how to open the setting panel, i.e., “Click the ‘Windows’ icon on the left side; Find the ‘Settings’ button and click it.” These writing styles can not only help our model to understand the steps but also help the readers with limited background on the technique area.

\section{Related Work}
\label{sec:related}

\subsection{Static Content Conversion}

Converting static materials (e.g., text, images) to multimedia formats (e.g., speech, slideshows, videos) can enhance the audiences' understanding. Thus, 
researchers have proposed to generate verbal descriptions~\cite{ramakrishnan2017non, ahmed2012accessible, zhang2017personalized, vtyurina2019verse, zhong2014justspeak} or automatic animations~\cite{xu2008animating, willett2018mixed}. T2V~\cite{hayashi2014t2v} creates TV-program-like animation from the text by using a TV program Making Language (TVML) converter. To leverage both images and text, Videolization~\cite{kalender2018videolization} and URL2Video~\cite{chi2020automatic} can convert web content to non-narrated videos containing a sequence of shots. To counter inadequate images in the manuals, Crosscast~\cite{xia2020crosscast} uses NLP to identify keywords and geographic information and automatically selects relevant photos.
Leake et al.~\cite{leake2020generating} identify concrete words from paragraphs to find relevant images online. HelpViz~\cite{zhong2021helpviz} uses the pre-trained model to extract the actions from text-based instructions. To enhance the interactive features, HowToCut~\cite{chi2021automatic} combines the step-by-step text from tutorials and visual instructions to provide interactive navigation in the instructional videos.
Existing approaches focus more on finding the corresponding visual materials for the given descriptive text; however, it is difficult for them to automatically emulate the actions described in the manual. We design a framework to automatically emulate the actions based on the procedural instructions and record the procedure as an instructional video. Our system prototype focuses on creating videos for software  configuration-related user manuals, and it could be extended to support other types of user manuals.  
% Emulating user behavior on the software and system is relatively convenient. 
% HelpViz~\cite{zhong2021helpviz} is the most similar solution to ours. Since HelpViz relies on using Android Debug Bridge (ADB) commands on the Android systems to manipulate the GUI elements remotely, it cannot be easily extended to other platforms. In contrast, we leverage the platform-independent computer vision technique, so our approach can support various platforms and systems.

\subsection{GUI Automation}

GUI automation is an underlying technology to emulate the user behaviors on the software by reproducing \emph{GUI events} on \emph{GUI elements} such as buttons, text boxes, selection lists, and icons~\cite{tuovenen2019mauto}. GUI events include dragging a mouse, clicking buttons, and applying various combinations of keystrokes. The software is triggered to perform desired actions by the GUI events. An emulator is responsible to produce the GUI events according to the received scripts. 

Invoking the APIs from the GUI element drivers is a common solution to manipulate the GUI elements. With the unique element ID or label, the emulator can precisely locate the corresponding GUI element and conduct the required action on it. The GUI element drivers are platform-dependent, e.g., desktop and web applications require different drivers. The typical supporting tools include Selenium WebDriver~\cite{holmes2006automating}, Microsoft WinAppDriver~\cite{winappdriver}, Appium~\cite{verma2017mobile}.

Besides GUI element drivers, computer vision (CV) techniques have been applied to achieve GUI automation. The CV-driven GUI automation can support almost all GUI-driven applications including image-based recognition, coordinate-based recognition, and optical character recognition (OCR), regardless of implementation, operating system, or even platform. The image-based recognition detects and determines the types and locations of GUI elements on the screen with a predefined element set. %Multiple open source or commercial tools are available in practice: Sikuli~\cite{yeh2009sikuli, chang2010gui}, JAutomate~\cite{alegroth2013jautomate}, and HP UFT~\cite{lalwani2013uft}.
The coordinate-based recognition used in Sikuli~\cite{yeh2009sikuli, chang2010gui} and JUnit~\cite{gomez2013reran} extracts coordinates on the screen to interact with the GUI elements. OCR recognizes the characters displayed on the screen and matches the text with the correct area to determine the relevant GUI elements. %Tesseract~\cite{smith2007overview} is a famous open-source OCR engine and has released its 5th version in 2021.
To make \TN{} platform-independent, we choose to use the CV-driven GUI automation technique. We combine the techniques mentioned above to identify and manipulate various types of GUI elements during manual processing and action emulation.

\section{Conclusion}
\label{sec:conclude}

This paper presents \TN{}, an automatic framework that converts software/system-related user manuals into instructional videos.
\TN{} normalizes various types of user manuals to Markdown format and extracts the action lists via natural language processing. 
According to the extracted actions, \TN{} emulates the user to manipulate the software/system in the emulation environment with the help of the computer vision technique. By recording the manipulation procedure, \TN{} generates the instructional videos. Through the studies with 10 participants and an online survey with 100 responses, we evaluate our automatically-generated videos and find that our system can effectively create instructional videos from software user manuals to help users understand and follow the software manipulation procedure.

% \section*{Acknowledgment}

% This work was partially supported by ONR grant N00014-18-2893.
%%-------------------------------------------------------
\bibliographystyle{ACM-Reference-Format}
\bibliography{reference}
%%-------------------------------------------------------
\end{document}